\documentclass[epj]{svjour}
\usepackage{amssymb,amsbsy,graphicx}
\usepackage{color}

\begin{document}

\title{Anomalous response in
the vicinity of spontaneous symmetry breaking}

\author{Seung Ki Baek\inst{1}
\and Hye Jin Park\inst{2}
\and Beom Jun Kim\inst{2}}

\institute{
   Department of Physics, Pukyong National University, Busan 608-737, Korea,
   \email{seungki@pknu.ac.kr}
   \and
   Department of Physics, Sungkyunkwan University, Suwon 440-746, Korea
   \email{beomjun@skku.edu}
}
\date{Received: date / Revised version: date}
\abstract{We propose a mechanism to induce negative AC permittivity in the
vicinity of a ferroelectric phase transition involved with
spontaneous symmetry breaking. This mechanism makes use of responses
at low frequency, yielding a high gain and a large phase delay, when the system
jumps over the free-energy barrier with the aid of external fields.
We illustrate the mechanism by analytically studying
spin models with the Glauber-typed dynamics under periodic perturbations.
Then, we show that the scenario is supported by numerical
simulations of mean-field as well as two-dimensional spin systems.
}

\maketitle

\section{Introduction}

When a many-body system is subjected to periodic driving, e.g., via
electromagnetic or acoustic waves, it has three different time scales:
One is a microscopic time scale associated with thermal fluctuations, and
we may regard this as a unit for measuring other time scales.
Another time scale is related to internal collective dynamics,
mediated by interatomic coupling. The other is the period of the
external driving.
In the absence of periodic driving, the system will relax to equilibrium,
and the fluctuation-dissipation relations provide a framework to relate
the macroscopic relaxation to microscopic dynamics~\cite{hanggi}.
If external driving is turned on, its period begins to compete with the
other time scales. The interplay of these time scales has been investigated
extensively in studies of stochastic resonance (see, e.g.,
Ref.~\cite{review} for a review). It is now well established that thermal
fluctuations can play a constructive role to enhance sensitivity to the
driving by modifying the internal time scale.

When a spin system is perturbed by an oscillating field coupled to
an order parameter, we have to observe the amplitude and phase of the order
parameter: In optics, for example, the optical response of an object to an
incident beam is determined by the interference between the beam and secondary
waves from the object, and their relative amplitudes and phase differences are
important factors in the interference phenomenon~\cite{hecht}.
By considering spin-spin correlation and its time scale
in the generation of the secondary waves,
we can obtain a more realistic picture of a real many-body system.
Interestingly, the mean-field (MF) spin dynamics,
the simplest description for collective ordering,
is similar to an overdamped oscillator, if written as an equation of motion for
total magnetization as will be explained below. For this reason,
the MF spin dynamics will serve as a starting point in our investigation.
Some progresses have been made through perturbative calculations
whereby the existence of stochastic resonance is predicted to the leading
order~\cite{leung,dsr,dsrq}.
One can also improve the prediction by taking
higher-order terms into account~\cite{adiabatic}.
However, we will point out a different kind of response that has been
missing in these perturbative approaches.
This motion is observed when the system hops from one free-energy minimum
to another (see, e.g., instantons~\cite{raj} for comparison).
It shows a large phase delay relative to the incident wave,
comparable to anomalous refraction. The frequency is nevertheless low.
Such a large delay would be impossible at low frequency if the system was
composed of weakly interacting simple harmonic oscillators.
Moreover, we argue that if the system is effectively described as
globally coupled, the response can be very strong compared with the input
signal. It is known that the linear response theory can break down at low
frequency~\cite{casado1,schmidt,alt}, and this failure has been related to
an anomalously high gain~\cite{casado2,casado3,casado4}. By considering
another important factor shaping the response, i.e., the phase difference,
we in this work will discuss anomalies in \emph{both} the amplitude and phase,
which we expect to open possibilities for interesting optical applications.

\section{Model}

Some materials at a sufficiently low temperature exhibit a phenomenon called
a polarization catastrophe, by which it acquires nonzero polarization $P$
under a vanishingly small external electric field $E$~\cite{kittel}.
The consequence is that the static permittivity, defined as
$\epsilon = \epsilon_0 + P/E$, diverges at this point, where
$\epsilon_0$ is the electric permittivity of free space.
In other words,
the system undergoes a ferroelectric phase transition as temperature $T$
varies. In the vicinity of the transition point $T=T_c$, the behavior of the
system can be phenomenologically described by the Landau
theory~\cite{kittel,primer}.
The central assumption is that the free-energy density can be
expanded as a polynomial in the order parameter $P$:
\begin{equation}
F_{\rm L} = -EP + g_0 + \frac{1}{2} \sigma (\beta_c-\beta) P^2 + \frac{1}{4} g_4
P^4 + \ldots,
\label{eq:landau}
\end{equation}
where $\sigma$ and $g_4$ are positive constants, and $\beta \equiv (k_B T)^{-1}$
is the inverse temperature with the Boltzmann constant $k_B$.  In terms of
$\beta$, the critical point is now expressed as $\beta_c \equiv (k_B T_c)^{-1}$.
Minimizing $F_{\rm L}$ with respect to $P$ requires
\begin{equation}
0 = -E + \sigma (\beta_c-\beta) P + g_4 P^3 + \ldots.
\end{equation}
When $E$ is absent, the theory predicts a continuous phase transition at
$\beta_c$ of the MF universality class.
Note that this is the universality class of three-dimensional quantum Ising
ferromagnets and uniaxial dipolar Ising
ferromagnets~\cite{guggen1,nielsen,guggen2}.

Let us investigate the dynamical aspects by
choosing the globally coupled Ising model
as a concrete example of the Landau theory.
We will discuss this model in magnetic terms as usual, but
it should be understood as covering electric systems as well,
if the magnetization order parameter is substituted by polarization $P$.
The Ising model
has actually been used to describe ferroelectric materials such as NaNO$_2$ or
(Glycine)$_3$·H$_2$SO$_4$~\cite{yamada1,yamada2,net,mitsui}.
One might point out that polarization varies continuously in a crystal,
making the naive Ising picture inappropriate.
However, the discreteness of an Ising spin
plays only a minor role and it is the up-down symmetry that is more
crucial~\cite{cardy}.
Let us consider the energy function of the Ising model,
\begin{equation}
U = -\frac{J}{N} \sum_{i>j} s_i s_j - h \sum_i s_i,
\label{eq:energy}
\end{equation}
where $J>0$ represents the ferromagnetic coupling and
$s_j$ is a binary Ising spin variable that
takes either of $\pm 1$ (see Ref.~\cite{vives} for a connection to the
antiferromagnetic case).
The response of the spin system to time-dependent $h$ is commonly studied by
using the Glauber dynamics~\cite{glauber}, which shows qualitative agreements
with experimental results~\cite{rmp}.  Under the Glauber dynamics, the
transition probability from spin configuration $\mu$ to $\nu$, differing by a
single spin, is defined as
\begin{equation}
w_{\nu\mu} = \frac{1}{1+\exp [-\beta (U_\mu - U_\nu)]},
\end{equation}
where $U_\eta$ means the energy of spin configuration $\eta = \{ s_1,
s_2, \ldots, s_N \}$ consisting of $N$ Ising spins, as defined in
equation~(\ref{eq:energy}).
By solving the master
equation, one can readily derive the effective free-energy density
functional as
\begin{equation}
\mathcal{F} = \frac{m^2}{2} - \beta^{-1} \ln \cosh \beta (m+h)
\label{eq:free}
\end{equation}
with magnetization $m = N^{-1} \sum_i s_i$ (see Ref.~\cite{dsrq} for the
details). It is important that we have already taken the thermodynamic
limit in equation~(\ref{eq:free}), so that this description is free from any
finite-size effects.
Thermal noises are also taken into account in this
description, because we work with the free-energy density functional.
This differs from a common approach through the time-dependent
Ginzburg-Landau phenomenology in the form of a nonlinear Langevin equation
(see, e.g., Refs.~\cite{casado1,schmidt}),
which contains a double-well potential and a noise term separately so
that every observable should be averaged over the noise.
We additionally note that only steady states are concerned
throughout this work, so that initial transients are assumed to have died
out in our analytic and numerical considerations.
In the context of ferroelectricity, the spin variable $s_i$ translates as an
electric dipole moment, while $m$ and $h$ correspond to $P$ and $E$,
respectively.
It is well known that the behavior of equation~(\ref{eq:free}) is qualitatively the
same as assumed in the Landau theory in equation~(\ref{eq:landau}).
The time evolution in the Glauber dynamics is formulated as
\begin{equation}
\frac{dm}{dt} = -\frac{\partial \mathcal{F}}{\partial m}
= -m + \tanh\beta (m+h).
\label{eq:dm}
\end{equation}
This is purely relaxational dynamics with no inertia, as indicated by the
absence of second-order derivatives, so it can be compared to an overdamped
oscillator.
In the limit of small $h \rightarrow 0$,
we may restrict ourselves to linear responses.
Then, the relaxation time $\tau$ is obtained as
\begin{equation}
\tau = \left[ 1 + \beta( {m^\ast}^2 - 1) \right]^{-1},
\label{eq:tau}
\end{equation}
where $m^\ast >0$ denotes the magnitude of magnetization in
equilibrium~\cite{dsrq}, which is nonzero only when $\beta > \beta_c = 1$.
If the free-energy functional for $\beta \gtrsim \beta_c$ and $h=0$
is approximated by
\begin{equation}
\mathcal{F} \approx \frac{\beta^3}{12} \left[ \left(m^2 - {m^\ast}^2
\right)^2 - {m^\ast}^4 \right],
\label{eq:quart}
\end{equation}
we readily find that ${m^\ast}^2 \approx 3\beta^{-3}(\beta-\beta_c)$
via the Taylor expansion.
This internal time scale $\tau$ is an important quantity
when a periodic external perturbation $h(t)$ with frequency
$f=(2\pi)^{-1}\omega$ is applied, because
stochastic resonance takes place
when the time-scale matching condition $f \approx \tau^{-1}$ is met to the
leading order~\cite{review,leung,dsr,dsrq,quantum}.
The minimum of $\mathcal{F}$ determines $m^\ast$, and
$\tau$ is related to the curvature around it.
We have one minimum when $\beta<\beta_c$,
while two minima appear when $\beta>\beta_c$ in the absence of $h$.
The important point is that each of them has a finite curvature, which allows
the system to relax to an equilibrium point within a finite time scale.
At $\beta=\beta_c$, on the other hand, the local curvature vanishes around
$m^\ast=0$, which is expressed by $\tau \rightarrow
\infty$ in the linear-response theory (see Eq.~(\ref{eq:tau})).

However, something is missing in this picture: When $\beta \gtrsim
\beta_c$, we can define an additional time scale
$\tau'$ involved with the \emph{swing} between $m^\ast$ and $-m^\ast$.
This swinging motion becomes possible when the energy
scale due to the periodic perturbation such as $h(t) = h_0 \cos \omega t$
overcomes the free-energy barrier
in the middle. But the energy scale needs not be large, because the barrier can
be made as small as we want by adjusting the temperature
close enough to the critical point.
If the external perturbation barely overcomes the barrier, the swinging
time scale $\tau'$ can be much larger than the usual
relaxation time $\tau$. It means
the existence of a critical field strength $h_c$ above which the swing becomes
possible, and $\tau'$ diverges as $h$ approaches $h_c$ from above. One can see
the reason by approximating equation~(\ref{eq:dm}) for small $\beta (m+h)$ as
follows:
\begin{equation}
\frac{dm}{dt} \approx (-m + \tanh\beta m) + \beta h,
\end{equation}
to the leading order in $\beta h \ll 1$. Let us
assume that the external field $h$
changes little while $m$ transits from $m^\ast$ to $-m^\ast$.
The expression inside the parentheses is an odd function of $m$, so one
may take the first sine Fourier component within an appropriate interval.
Here, we simply observe that it vanishes at $\pm m^\ast$ and that
its slope at $m=0$ equals $(\beta - \beta_c) m$ to approximate
the functional shape as
\begin{equation}
\frac{dm}{dt} \approx \frac{(\beta-\beta_c) m^\ast}{\pi}
\sin\left(\frac{m}{m^\ast}\pi\right) + \beta h.
\label{eq:approx}
\end{equation}
It is noteworthy that equation~(\ref{eq:approx}) can also be motivated by
considering a \emph{gravity pendulum} in the overdamped
limit~\cite{huber,gwinn,grebogi,jj}. The gravity pendulum should be
inverted upside down to describe one unstable fixed point ($m=0$) and two
stable fixed points ($\pm m^\ast$).
The swing time $\tau'$ can be estimated from equation~(\ref{eq:approx}) as
\begin{equation}
\tau' \approx \int_{-m^\ast}^{+m^\ast} \frac{dt}{dm} dm
\approx \frac{2m^\ast}{\beta \sqrt{h^2-h_c^2}},
\label{eq:intermit}
\end{equation}
where $h_c \equiv \pi^{-1} m^\ast (1-\beta_c/\beta) \propto
(T_c-T)^{3/2}$.
Clearly, the divergence of $\tau'$ as $h \rightarrow h_c^+$ has the same origin
as the type-I intermittency~\cite{pm}.

Let us now take the temporal variations of the external field into
account. We will ignore the high-order terms of $m$ in
equation~(\ref{eq:dm}), because those terms are visible only when
$|m|$ is sufficiently large, whereas we have argued above that
the system spends most of the time in crossing over the barrier at small $|m|$.
Based on this argument, we consider the following differential equation:
\begin{equation}
\frac{dm}{dt} = (\beta - \beta_c) m + \beta h_0 \cos \omega t,
\label{eq:smallm}
\end{equation}
where $\beta \gtrsim \beta_c$. One obtains a physically
acceptable solution as
\begin{equation}
m(t) = \left[ \frac{\beta}{\sqrt{(\beta-\beta_c)^2 + \omega^2}} \right] h_0
\cos(\omega t + \phi + \pi),
\label{eq:lin}
\end{equation}
where
\begin{equation}
\phi \equiv \arctan \left( \frac{\omega}{\beta-\beta_c} \right).
\end{equation}
Discarding other nonphysical solutions which blow up to infinity,
we are actually relying on the existence of high-order terms, which
seem to be neglected in equation~(\ref{eq:smallm}).

Even this crude picture hints at some of interesting characteristics.
First of all, the phase delay $\delta = \phi + \pi$
lies between $\pi$ and $\frac{3}{2}\pi$ so that
the overall sign of $m(t)$ is opposite to
that of $h(t)$. This simply means that
$h(t)$ should be positive to drive the system
with $m(t)<0$ to the other side, and vice versa. However, such a
large phase delay would not be seen if we regarded this system as an
overdamped oscillator inside a well, whose phase delay is bounded from above by
$\frac{\pi}{2}$~\cite{dsrq}.
Note also that
we can practically quantify the phase delay $\delta$ by calculating
\begin{equation}
\tan\delta = \frac{\chi''}{\chi'},
\label{eq:delay}
\end{equation}
where we have used Fourier cosine and sine components defined as
\begin{equation}
\begin{array}{l}
\chi' = \frac{1}{\pi h_0} \int_0^{2\pi} m(t) \cos \omega t~d(\omega t),\\
\chi'' = \frac{1}{\pi h_0} \int_0^{2\pi} m(t) \sin \omega t~d(\omega t).
\end{array}
\label{eq:acsus}
\end{equation}
Let us recap this characteristic in terms of $\tau'$:
It takes $2 \tau'$ to make a full swing back and forth,
and this should be shorter than the driving period $f^{-1}$ to sustain
the motion. Otherwise, the field will change the direction before the system
overcomes the barrier.
This can be understood as the time-scale matching condition for this mode.
From this consideration, we derive the following inequality:
\begin{equation}
2\tau' < f^{-1}.
\end{equation}
At the same time, if the system follows the field too quickly, the
phase difference will not be as large as suggested above. We are interested
in a condition to have $m<0$ for a sufficiently long time when $h>0$,
and vice versa. We already know that the field will be positive
for a half period $(2f)^{-1}$, and require that the duration of $m<0$
should occupy more than a half of it, i.e., $(4f)^{-1}$.
We thus arrive at the following inequality:
\begin{equation}
(4f)^{-1} < \int_{-m^\ast}^{0} \frac{dt}{dm} dm.
\end{equation}
When $h$ is close to $h_c$, the right-hand side (RHS) is comparable to $\tau'$
(Eq.~(\ref{eq:intermit})),
because the system spends most time in jumping over the free-energy barrier at
$m<0$. After some algebra, one can summarize these two conditions as follows:
\begin{equation}
f_c < f < 2f_c,
\label{eq:bnd}
\end{equation}
where $f_c \equiv (8\pi)^{-1} \sqrt{(\pi \beta h / m^\ast)^2
- (\beta-\beta_c)^2}$, and $m^\ast = \sqrt{3\beta^{-3} (\beta - \beta_c)}$.
In addition, noting that the calculation of $\tau'$ in
equation~(\ref{eq:intermit})
approximates a sinusoidal driving to a constant field, we may identify
$h^2$ in equation~(\ref{eq:bnd}) with a squared average, $\int_0^{2\pi}
h_0^2 \cos^2 (\omega t) d(\omega t) = h_0^2/2$.

The discussion above tells us that this mode is possible at low
frequency, if $h \rightarrow h_c^+$ (see Eqs.~(\ref{eq:intermit}) and
(\ref{eq:bnd})). In this respect, we can ignore $\omega^2$ in
equation~(\ref{eq:lin})
compared with $(\beta-\beta_c)^2$, which is kept constant, and
the expression inside the square brackets in equation~(\ref{eq:lin}) can
then be greater than $O(1)$. We argue the reason as follows:
As soon as we bring the system from one minimum to the top of the
free-energy barrier with $h \gtrsim h_c$, the system spontaneously moves over
to the other side. During this process, therefore,
the traveling distance can be very large
compared with $h$, because $h_c$ is an arbitrarily small quantity and the
free-energy landscape is quite flat when $\beta \gtrsim \beta_c$.
It implies a possibility of high gain in the sense that the resulting change
of the order parameter can be large with respect to the input field strength.
Of course, it should be noted that this amplification factor is
system-dependent, while the negative response to $h(t)$ is universally
expected.

A low-dimensional system may be an example of such system dependence.
In a finite-dimensional space, spatial variations of $m$
over coordinates $\mathbf{r}$
also contribute to the free energy in such a way that
\begin{equation}
\mathcal{F}[m(\mathbf{r})] = \int \left\{ \frac{\left[ \nabla
m(\mathbf{r}) \right]^2}{2} + V_\beta[m(\mathbf{r})] \right\} d\mathbf{r},
\label{eq:lowdim}
\end{equation}
where $V_\beta$ can be identified with the RHS of equation~(\ref{eq:free}).
By assuming relaxational dynamics to minimize equation~(\ref{eq:lowdim})
with respect to $m(\mathbf{r})$, one derives a minimal description with
a Laplacian term added to equation~(\ref{eq:dm})~\cite{krap}:
\begin{equation}
\frac{\partial m(\vec{r},t)}{\partial t} = \nabla^2 m - m + \tanh \beta(m+h).
\end{equation}
This description assumes that the system lies deep enough inside the ordered
phase to neglect critical-point fluctuations for the sake of
self-consistency~\cite{cardy}.
Put in the momentum space, the equation transforms to
\begin{equation}
\frac{\partial m(\vec{k},t)}{\partial t} = -( k^2+1) m + \tanh \beta(m+h),
\label{eq:kspace}
\end{equation}
where $k \equiv |\vec{k}|$.
We expect a characteristic length scale to exist in this system
because $\beta > \beta_c$.
If that is the case, we may consider only this characteristic mode of nonzero
$\vec{k}$, which would correspond to the typical size of spin domains.
Then, the amplification factor is estimated by the crude calculation above
(see Eq.~(\ref{eq:lin}))
as $\beta/ \sqrt{(\beta-\beta_c-k^2)^2 + \omega^2}$, which cannot
grow indefinitely due to the nonzero $k$, even if
$\beta$ is close to $\beta_c$ and $\omega$ is small.
This argument suggests that the phenomenon reported here can be diminished in
magnitude by domain dynamics that incorporates nontrivial spin fluctuations
over space. For this reason, we speculate that the response of a
low-dimensional system will not be as strong as in the globally coupled case.

\section{Numerical results}

\begin{figure}
\includegraphics[width=0.45\textwidth]{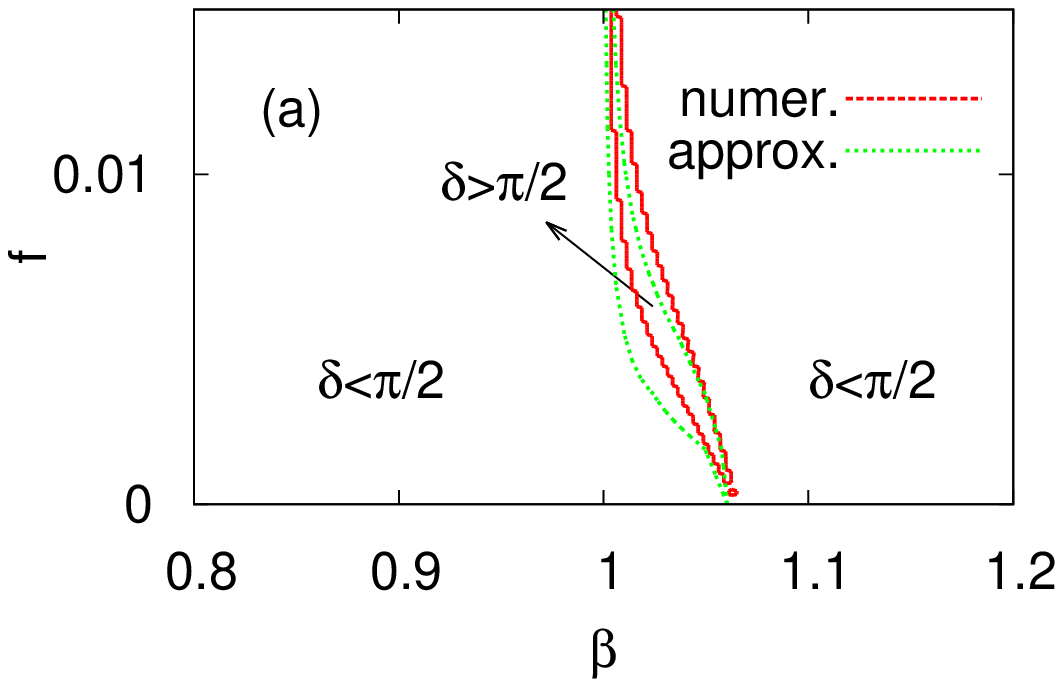}
\includegraphics[width=0.45\textwidth]{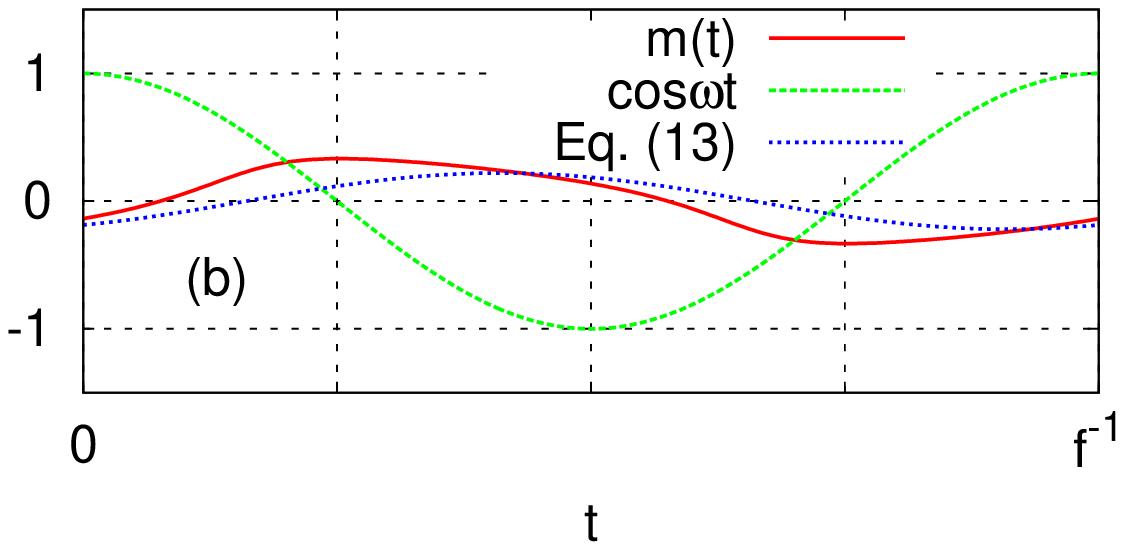}
\caption{(a) Phase diagram of the MF kinetic Glauber-Ising model,
driven by $h=h_0 \cos\omega t$ with $h_0 =
10^{-2}$. We have numerically integrated equation~(\ref{eq:dm}) by using
Heun's method with time-step size $\Delta t = 10^{-3}$ to obtain the solid
lines.
The phase delay $\delta$ is estimated from equation~(\ref{eq:delay}).
The dotted lines are obtained by substituting
$h^2 = h_0^2 / 2$ into equation~(\ref{eq:bnd}).
(b) Time series of $m$ responding oppositely to the external perturbation,
when $\beta = 1.04$ in units of $J^{-1}$,
and $f=4\times 10^{-3}$. We use the same $h_0$ and $\Delta t$ as in panel (a).
The solid line is obtained by integrating equation~(\ref{eq:dm}).
For comparison, the other curves show $\cos \omega t$ and the approximate
expression in equation~(\ref{eq:lin}), respectively.
Note that the amplitude is much larger than $h_0 = 10^{-2}$.
One can also check from equation~(\ref{eq:acsus}) that $\chi'$ is negative
whereas $\chi''$ is not, yielding $\delta > \frac{\pi}{2}$.
}
\label{fig:ising}
\end{figure}

\begin{figure}
\includegraphics[width=0.45\textwidth]{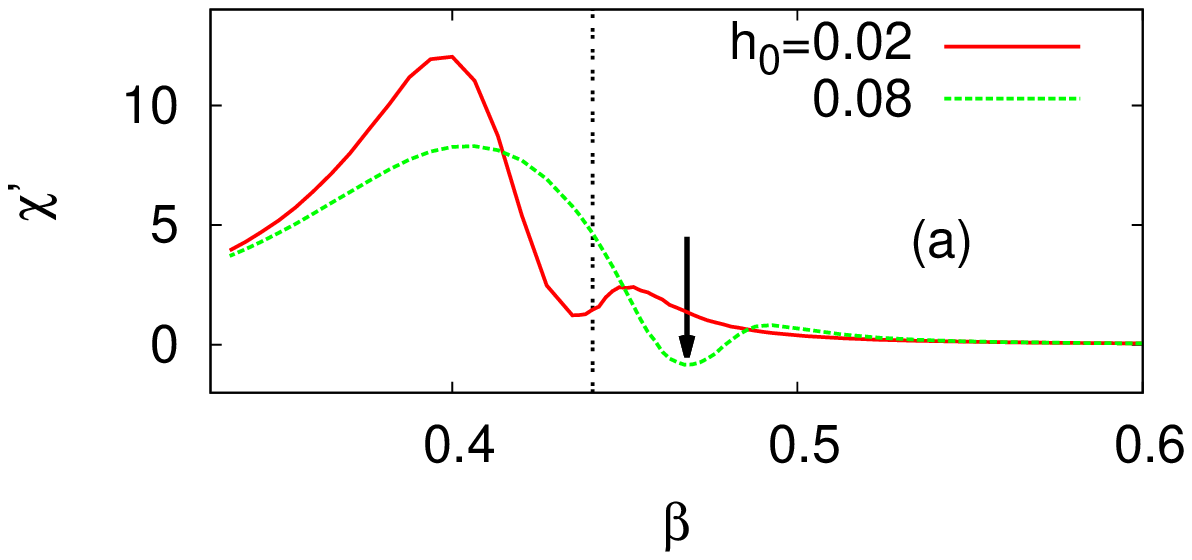}
\includegraphics[width=0.45\textwidth]{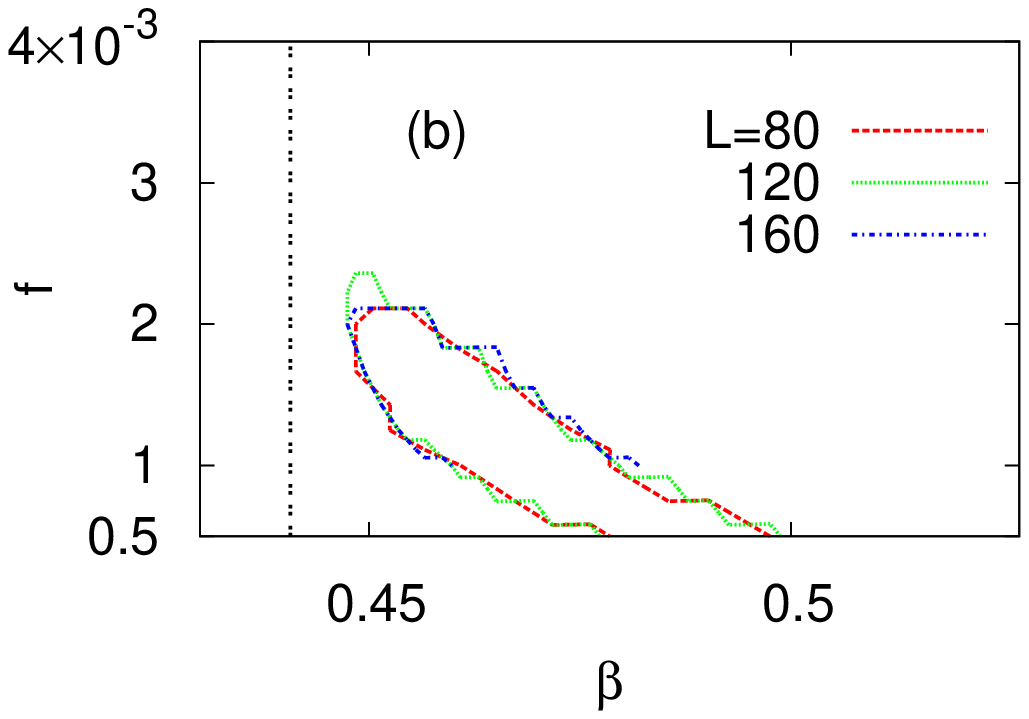}
\caption{Response of 2D Glauber-Ising system
driven by a periodic external field with amplitude $h_0$
and frequency $f= 10^{-3}$, as a function of $\beta$.
These plots are obtained by simulating Glauber-typed Monte Carlo dynamics on
an $L\times L$ square lattice with periodic boundary conditions.
The vertical dotted lines represent the 2D critical point $\beta_c^{(2D)} =
\ln(1+\sqrt{2})/2 \approx 0.44$.
(a) The negative dip in $\chi'$, indicated by the arrow, is found
above $\beta_c^{\rm (2D)}$. In this plot, the system size is chosen as $L=80$.
(b) We observe a region in which phase delay $\delta$ lies outside $[0,\pi/2]$
when $h_0=0.08$.
The curves look like zigzags because we have sampled grid
points on the $(\beta,f)$ plane with finite resolution.
}
\label{fig:2d}
\end{figure}

The characteristic large phase delay
is indeed observed in our numerical integration of the MF dynamics,
equation~(\ref{eq:dm}). Plotting $\delta$ in the $(\beta,f)$ plane,
we find such a region inside the ordered phase at low frequency
where $\delta$ exceeds $\frac{\pi}{2}$
(Fig.~\ref{fig:ising}a).
Note that equation~(\ref{eq:bnd}) qualitatively explains the
shape of the region.
Moreover, the response amplifies the input field
with a high gain by an order of magnitude inside the region
(Fig.~\ref{fig:ising}b).
If further lowering $f$, we observe a discontinuous phase transition which is
explained in the adiabatic approximation~\cite{adiabatic}.

Let us check how the behavior is affected by low dimensionality.
The two-dimensional (2D) Glauber-Ising model has an energy function
\begin{equation}
U = -J \sum_{\left< i,j \right>} s_i s_j - h \sum_i s_i,
\end{equation}
where  $\sum_{\left<i,j\right>}$ runs over the nearest neighbors.
It can also yield an inverted signal,
but the amplitude is not so large as in the MF model (Fig.~\ref{fig:2d}a),
as was explained by equation~(\ref{eq:kspace}) above. Whereas the region of
$\delta > \pi/2$ extends to high $f$ in the MF Ising model (see
Fig.~\ref{fig:ising}a), it is not detected at $f \gtrsim 2\times 10^4$
in our numerical calculation of the 2D model (Fig.~\ref{fig:2d}b).
Still, an important point is that the region of large phase delays survives
when we increase the system size.

\begin{figure}
\includegraphics[width=0.45\textwidth]{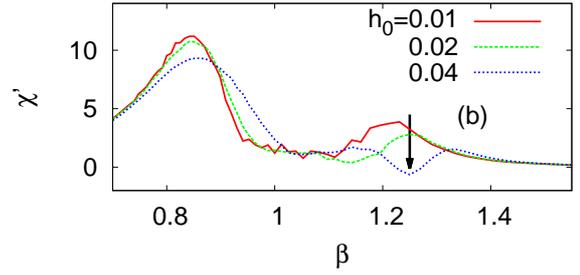}
\caption{Response of 2D five-state clock model
driven by a periodic external field with amplitude $h_0$
and frequency $f= 10^{-3}$ in the perpendicular direction to the magnetization.
We plot $\chi'$ as a function of $\beta$
by simulating Glauber-typed dynamics on
an $80\times 80$ square lattice with periodic boundary conditions.
In equilibrium, the model experiences spontaneous symmetry breaking around
$\beta \approx 1.1$~\cite{bori}, above which the negative dip is located
(the arrow).
}
\label{fig:q5}
\end{figure}

As another example, we consider the 2D five-state clock
model~\cite{jkkn,elit} on an $L \times L$ square lattice with size $N = L^2$.
Its energy function is given as
\begin{equation}
E = -J \sum_{\left<i,j\right>} \mathbf{s}_i \cdot \mathbf{s}_j - \mathbf{h}
\cdot \sum_{i} \mathbf{s}_i,
\end{equation}
where each spin $\mathbf{s}_j = (\cos\theta_j, \sin\theta_j)$ at
site $j=1,\ldots,N$ has a discrete angle $\theta_j = 2\pi n_j/q$
with $n_j = 0, 1, \ldots, 4$. The magnetization of the system is defined as a
vector sum $\mathbf{m} = N^{-1} \sum_j \mathbf{s}_j$. A weak and slow driving
in-plane field $\mathbf{h}$ is applied in a perpendicular direction to
$\mathbf{m}$
with amplitude $h_0 \ll 1$ and angular frequency $\omega \ll 1$~\cite{dsr2d}.
The model in equilibrium undergoes double phase transitions, only one of which
at the lower $T$ is accompanied by spontaneous symmetry breaking. Although its
critical properties belong to a different universality class from that of the
2D Ising model~\cite{q5}, Fig.~\ref{fig:q5} shows that the symmetry
breaking can induce a response with a large phase delay, manifested by a
negative dip in $\chi'$, as long as the field supplies enough driving force to
overcome the small free-energy barrier.
This is consistent with our understanding that the anomalous behavior is not
specific to particular model systems but related to general features of the
symmetry-breaking phenomenon.

\section{Discussion and summary}

Although the existence of anomalous response with a large phase delay
is precluded in the overdamped limit of the linear-response theory,
we have found that it still remains possible as a non-perturbative mode.
We have also argued the characteristic features
in the general terms of free-energy landscapes, meaning that it is quite a
general mechanism involved with spontaneous symmetry breaking itself.
The magnitude of the response depends on which specific system we are dealing
with, and we expect the effect to be better manifested if the system
fits better into MF-like dynamics.
In particular, if polarization $P$ is opposite to the external field $E$
with sufficiently large magnitude, it implies that the AC permittivity
$\epsilon (\omega) = \epsilon_0 + P/E$ may become negative.
Whether this effect is observable experimentally
remains to be tested in a future study.

\begin{acknowledgement}
S.K.B. was supported by Basic Science Research Program through the
National Research Foundation of Korea (NRF) funded by the Ministry of Science,
ICT and Future Planning (NRF-2014R1A1A1003304).
B.J.K. was supported by the National Research Foundation of Korea (NRF) grant
funded by the Korea government (MSIP) (No. NRF-2014R1A2A2A01004919).
\end{acknowledgement}


\begin{thebibliography}{39}

\bibitem{hanggi}
P.~H{\"a}nggi, H.~Thomas, Phys. Rep. \textbf{88}, 207 (1982)

\bibitem{review}
L.~Gammaitoni, P.~H{\"a}nggi, P.~Jung, F.Marchesoni, Rev. Mod. Phys.
  \textbf{70}, 223 (1998)

\bibitem{hecht}
E.~Hecht, \emph{Optics}, 4th~edn. (Addison Wesley, San Francisco, 2002)

\bibitem{leung}
K.T. Leung, Z.~N{\'e}da, Phys. Lett. A \textbf{246}, 505 (1998)

\bibitem{dsr}
B.J. Kim, P.~Minnhagen, H.J. Kim, M.Y. Choi, G.S. Jeon, EPL \textbf{56}, 333
  (2001)

\bibitem{dsrq}
S.K. Baek, B.J. Kim, Phys. Rev. E \textbf{86}, 011132 (2012)

\bibitem{adiabatic}
B.J. Kim, H.~Hong, J. Korean Phys. Soc. \textbf{52}, 203 (2008)

\bibitem{raj}
R.~Rajaraman, \emph{Solitons and Instantons} (North Holland, Amsterdam, 1987)

\bibitem{casado1}
J.~Casado-Pascual, J.~G{\'o}mez-Ord{\'o}{\~n}ez, M.~Morillo, P.~H{\"a}nggi, EPL
  \textbf{58}, 342 (2002)

\bibitem{schmidt}
L.~Schmidt, R.R. Netz, EPL \textbf{98}, 10014 (2012)

\bibitem{alt}
S.K. Baek, F.~Marchesoni, Phys. Rev. E \textbf{89}, 022136 (2014)

\bibitem{casado2}
J.~Casado-Pascual, J.~G{\'o}mez-Ord{\'o}{\~n}ez, M.~Morillo, P.~H{\"a}nggi,
  Phys. Rev. Lett. \textbf{91}, 210601 (2003)

\bibitem{casado3}
J.~Casado-Pascual, C.~Denk, J.~G{\'o}mez-Ord{\'o}{\~n}ez, M.~Morillo,
  P.~H{\"a}nggi, Phys. Rev. E \textbf{67}, 036109 (2003)

\bibitem{casado4}
J.~Casado-Pascual, J.~G{\'o}mez-Ord{\'o}{\~n}ez, M.~Morillo, P.~H{\"a}nggi,
  Phys. Rev. E \textbf{68}, 061104 (2003)

\bibitem{kittel}
C.~Kittel, \emph{Introduction to Solid State Physics} (John Wiley \& Sons, New
  York, 1953)

\bibitem{primer}
P.~Chandra, P.~Littlewood, in \emph{Physics of Ferroelectrics} (Springer,
  Berlin, 2007), Vol. 105 of \emph{Topics in Applied Physics}, pp. 69--116

\bibitem{guggen1}
N.J. Als-Nielsen, L.~Holmes, H.~Guggenheim, Phys. Rev. Lett. \textbf{32}, 610
  (1974)

\bibitem{nielsen}
N.J. Als-Nielsen, Phys. Rev. Lett. \textbf{37}, 1161 (1976)

\bibitem{guggen2}
G.~Ahlers, A.~Kornblit, H.J. Guggenheim, Phys. Rev. Lett. \textbf{34}, 1227
  (1975)

\bibitem{yamada1}
Y.~Yamada, I.~Shibuya, S.~Hoshino, J. Phys. Soc. Jpn. \textbf{18}, 1594 (1963)

\bibitem{yamada2}
Y.~Yamada, Y.~Fujii, I.~Hatta, J. Phys. Soc. Jpn. \textbf{24}, 1053 (1968)

\bibitem{net}
R.E. Nettleton, Ferroelectrics \textbf{2}, 77 (1971)

\bibitem{mitsui}
T.~Mitsui, E.~Nakamura, M.~Tokunaga, Ferroelectrics \textbf{5}, 185 (1973)

\bibitem{cardy}
J.~Cardy, \emph{Scaling and Renormalization in Statistical Physics} (Cambridge
  University Press, Cambridge, 2002)

\bibitem{vives}
E.~Vives, T.~Cast\'{a}n, A.~Planes, Am. J. Phys. \textbf{65}, 907 (1997)

\bibitem{glauber}
R.J. Glauber, J. Math. Phys. \textbf{4}, 294 (1963)

\bibitem{rmp}
B.K. Chakrabarti, M.~Acharyya, Rev. Mod. Phys. \textbf{71}, 847 (1999)

\bibitem{quantum}
S.G. Han, J.~Um, B.J. Kim, Phys. Rev. E \textbf{86}, 021119 (2012)

\bibitem{huber}
B.A. Huberman, J.P. Crutchfield, N.H. Packard, Appl. Phys. Lett. \textbf{37},
  750 (1980)

\bibitem{gwinn}
E.G. Gwinn, R.M. Westervelt, Phys. Rev. Lett. \textbf{54}, 1613 (1985)

\bibitem{grebogi}
C.~Grebogi, E.~Ott, J.A. Yorke, Phys. Rev. Lett. \textbf{56}, 1011 (1986)

\bibitem{jj}
B.J. Kim, P.~Minnhagen, P.~Olsson, Phys. Rev. B \textbf{59}, 11506 (1999)

\bibitem{pm}
Y.~Pomeau, P.~Manneville, Commun. Math. Phys. \textbf{74}, 189 (1980)

\bibitem{krap}
P.L. Krapivskiy, S.~Redner, E.~Ben-Naim, \emph{A Kinetic View of Statistical
  Physics} (Cambridge University Press, Cambridge, 2010)

\bibitem{jkkn}
J.V. Jos{\'e}, L.P. Kadanoff, S.~Kirkpatrick, D.R. Nelson, Phys. Rev. B
  \textbf{16}, 1217 (1977)

\bibitem{elit}
S.~Elitzur, R.B. Pearson, J.~Shigemitsu, Phys. Rev. D \textbf{19}, 3698 (1979)

\bibitem{dsr2d}
H.J. Park, S.K. Baek, B.J. Kim, Phys. Rev. E \textbf{89}, 032137 (2014)

\bibitem{q5}
S.K. Baek, H.~M{\"a}kel{\"a}, P.~Minnhagen, B.J. Kim, Phys. Rev. E \textbf{88},
  012125 (2013)

\bibitem{bori}
O.~Borisenko, G.~Cortese, R.~Fiore, M.~Gravina, A.~Papa, Phys. Rev. E
  \textbf{83}, 041120 (2011)

\end{thebibliography}

\end{document}